\documentclass[%
 reprint,
 superscriptaddress,
 nofootinbib,
 amsmath,amssymb,
 prc,
 floatfix
]{revtex4-2}

\usepackage{graphicx}% Include figure files
\usepackage{dcolumn}% Align table columns on decimal point
\usepackage{bm}% bold math
\usepackage{xcolor}
\usepackage{wasysym}
\usepackage{physics}
\usepackage{amsmath}
\usepackage[pdftex,colorlinks,linkcolor = blue,urlcolor  = blue,citecolor = blue]{hyperref}
\begin{document}

\preprint{APS/123-QED}

\title{Nuclear level densities and $\gamma-$ray strength functions of $^{111,112,113}$Sn isotopes studied with the Oslo method}% Force line breaks with \\

\author{M.~Markova}
\email{maria.markova@fys.uio.no}
\affiliation{Department of Physics, University of Oslo, N-0316 Oslo, Norway}

\author{A.~C.~Larsen}
\email{a.c.larsen@fys.uio.no}
\affiliation{Department of Physics, University of Oslo, N-0316 Oslo, Norway}

%\author{P.~von Neumann-Cosel}%
%\affiliation{%
%Institut f\"{u}r Kernphysik, Technische Universit\"{a}t Darmstadt, D-64289 Darmstadt, Germany
%}%

\author{G.~M.~Tveten}
\affiliation{Department of Physics, University of Oslo, N-0316 Oslo, Norway}
\affiliation{Expert Analytics AS, N-0179 Oslo, Norway}

\author{P.~von Neumann-Cosel}%
\affiliation{%
 Institut f\"{u}r Kernphysik, Technische Universit\"{a}t Darmstadt, D-64289 Darmstadt, Germany
}%

\author{T.~K.~Eriksen}
\affiliation{Department of Physics, University of Oslo, N-0316 Oslo, Norway}

\author{F.~L.~Bello Garrote}
\affiliation{Department of Physics, University of Oslo, N-0316 Oslo, Norway}

\author{L.~Crespo Campo}
\affiliation{Department of Physics, University of Oslo, N-0316 Oslo, Norway}

\author{F.~Giacoppo}
\affiliation{GSI Helmholtzzentrum für Schwerionenforschung, Planckstraße 1, 64291 Darmstadt, Germany}
%\affiliation{Helmholtz-Institut Mainz, Staudingerweg 18, 55128 Mainz, Germany}

\author{A.~G\"{o}rgen}
\affiliation{Department of Physics, University of Oslo, N-0316 Oslo, Norway}

\author{M.~Guttormsen}
\affiliation{Department of Physics, University of Oslo, N-0316 Oslo, Norway}

\author{K.~Hadynska-Klek}
\affiliation{Heavy Ion Laboratory, University of Warsaw, Ludwika Pasteura 5A, 05-077 Warszawa,
Poland}

\author{M.~Klintefjord}
\affiliation{Department of Physics, University of Oslo, N-0316 Oslo, Norway}

%\author{H.~T.~Nyhus}
%\affiliation{Department of Physics, University of Oslo, N-0316 Oslo, Norway}
%\affiliation{Expert Analytics AS, N-0179 Oslo, Norway}

\author{T.~Renstr{\o}m}
\affiliation{Department of Physics, University of Oslo, N-0316 Oslo, Norway}
\affiliation{Expert Analytics AS, N-0179 Oslo, Norway}

\author{E.~Sahin}
\affiliation{Department of Physics, University of Oslo, N-0316 Oslo, Norway}

\author{S.~Siem}
\affiliation{Department of Physics, University of Oslo, N-0316 Oslo, Norway}

\author{T.~G.~Tornyi}
\affiliation{Department of Physics, University of Oslo, N-0316 Oslo, Norway}

\date{\today}% It is always \today, today,

\begin{abstract}
The $^{111,112,113}$Sn isotopes have been studied with ($p,d \gamma$), ($p,p^{\prime} \gamma$), and ($d,p \gamma$) reactions to extract the nuclear level densities (NLDs) and $\gamma$-ray strength functions (GSFs) of these nuclei below the neutron separation energy by means of the Oslo method. The experimental NLDs for all three nuclei demonstrate a trend compatible with the constant-temperature model below the neutron separation energy while also being in good agreement with the NLDs of neighboring Sn isotopes, obtained previously with the Oslo-type and neutron evaporation experiments. The extracted microcanonical entropies yield $\approx 1.5$ $k_B$ entropy of a valence neutron in both $^{111}$Sn and $^{113}$Sn. Moreover, the deduced microcanonical temperatures indeed suggest a clear constant-temperature behavior above $\approx$ 3 MeV in $^{111,113}$Sn and above $\approx$ 4.5 MeV in $^{112}$Sn. We observe signatures for the first broken neutron pairs between 2 and 4 MeV in all three nuclei. 
The GSFs obtained with the Oslo method are found to be in good agreement below the neutron threshold with the strengths of $^{112,114}$Sn extracted in the ($p,p^{\prime}$) Coulomb excitation experiments.
\end{abstract}

\maketitle
%---------------------------- Section 1: INTRODUCTION ---------------------------
\section{\label{sec 1: introduction}Introduction}

The statistical approach to the description of excited nuclei has always been an integral part of reaction theory since its first introduction and application in 1952 by Hauser and Feshbach \cite{Hauser1952}. 
This remains true today, and the statistical model has grown into an indispensable tool for modelling  nuclear reactions for astrophysics \cite{Arnould2007}, reactor design and waste transmutation \cite{Chadwick2011,Salvatores2011}, and medical isotope production \cite{Rahman2020}. Two key inputs needed for the statistical-model calculations are the nuclear level density (NLD) and, in case of reactions involving photons, the $\gamma$-ray strength function (GSF). 
The NLD, $\rho(E_x)$, provides a measure of a number of quantum mechanical levels available at a given excitation energy $E_x$, whereas the GSF, $f(E_{\gamma})$, characterizes an average, reduced $\gamma$-transition probability as a function of $\gamma$-ray energy $E_{\gamma}$. 
Besides their importance for reaction cross sections and rate estimations, both of these average nuclear characteristics provide a critical insight into nuclei as complex many-body systems, their structure and decay properties in the quasi-continuum and continuum excitation energy regimes. 

At relatively low excitation energies, within the discrete region, the NLD can be straightforwardly found through counting known discrete levels (e.g. available in compilations such as provided in Ref.~\cite{ensdf}) with conventional spectroscopy. 
After the onset of  Cooper-pair breaking  at higher excitation energies, the NLD increases exponentially, and experimental spectroscopic data tend to underestimate its values drastically. 
In this energy range, the experimental information on NLDs can be obtained from, for example, particle evaporation spectra \cite{Vonach1966} or by a fluctuation analysis of fine structures of giant resonances studied in high-energy light-ion reactions at extreme forward angles \cite{Martin2017,Kalmykov07}. 
Nuclear resonance fluorescence, inelastic relativistic proton scattering, discrete resonance capture and other experimental techniques reviewed in detail in Ref.~\cite{Goriely2019a} provide an access to the GSFs below and above the neutron separation energy. In this work, we make use of the Oslo method \cite{Guttormsen1987,Guttormsen1996,Schiller2000}, an experimental technique where the NLD and GSF are simultaneously extracted for excitation energies below the neutron threshold. 
This method has been %repeatedly shown to produce results in general good agreement with other methods, further 
used for addressing numerous key questions, such as the validity of the Brink-Axel hypothesis \cite{Guttormsen16, Campo2018}, study of thermal properties of excited nuclei \cite{Agvaanluvsan09-1,Nyhus2012}, constraining the radiative neutron capture cross sections relevant for astrophysical $s$ and $r$ processes \cite{Kullmann2019,Spyrou2014}\footnote{The latter Ref. exploits the Oslo method combined with $\beta$-decay measurements, or the so-called $\beta$-Oslo method.}, and more.

From the perspective of investigating statistical properties, Sn isotopes provide us with excellent study cases, where the Oslo-method NLDs and GSFs can be directly compared to numerous experimental results and theoretical predictions. 
Moreover, it becomes possible to combine these cases in a broader systematic study of statistical properties of nuclei with an increasing neutron number performed with the same method. 
At the moment, the NLDs and GSFs have been reported for the Oslo-type studies of $^{116,117}$Sn \cite{Agvaanluvsan09-1, Agvaanluvsan2009-2}, $^{118,119}$Sn \cite{Toft2010}, $^{121,122}$Sn \cite{toft2011}, and $^{120,124}$Sn \cite{Markova2022}. 
The measurements for the latter two isotopes were performed with the new scintillator detector array OSCAR \cite{Ingeberg2020, Zeiser2020}, currently available at the Oslo Cyclotron Laboratory (OCL).

Additional studies with the Oslo method on lighter Sn isotopes are highly desired to complete this sort of a systematic review. 
Constraining the statistical properties and putting them into the context of systematics for neutron-deficient Sn isotopes might also be of further interest to shed new light on the rapid proton-capture process that can take place on accreting neutron stars (e.g., Ref.~\cite{Schatz2001}). This work presents the NLDs and GSFs for $^{111,112,113}$Sn nuclei obtained from particle-$\gamma$ coincidence data by means of the Oslo method. In Sec.~\ref{sec 2: Experiment} we present the details regarding the experimental setup at the OCL and some of the most important steps of the data processing. Section \ref{sec 3: Oslo method}  covers the details of the Oslo-method implementation for the extraction of NLDs (Subsec.~\ref{subsec 3.1: NLDs}) and GSFs (Subsec.~\ref{subsec 3.2: GSFs}). 
The main results on the NLDs in $^{111,112,113}$Sn and their thermal properties as well as the GSFs are presented in Sec.~\ref{sec 4: NLD+Thermo} and Sec.~\ref{sec 5: GSF}, respectively. Finally, the main conclusions are outlined in Sec.~\ref{sec 6: Conclusion}.                                                                                                                                                                     
%---------------------------- Section 2: EXPERIMENT -----------------------------
\section{\label{sec 2: Experiment}Experimental setup and data processing}

The $^{111,112,113}$Sn isotopes were studied at the OCL in $^{112}$Sn($p,p^{\prime}\gamma$)$^{112}$Sn, $^{112}$Sn($p,d\gamma$)$^{111}$Sn, and $^{112}$Sn($d,p\gamma$)$^{113}$Sn reactions performed on a self-supporting 99.8\% enriched $^{112}$Sn foil target of 4 mg/cm$^2$ thickness. 
Proton beams with energies of 25 MeV and 16 MeV provided by the MC-35 Scanditronix cyclotron were used to investigate $^{111}$Sn and $^{112}$Sn in ($p,p^{\prime}\gamma$) and ($p,d\gamma$) reactions, respectively. Beam intensities were kept at $I \approx 1.0-1.5$ nA in both cases.  The $^{113}$Sn nucleus was studied with a 11.5 MeV deuteron beam with intensities of $I \approx 0.5-0.7$ nA. 

%------------------------------------ FIG_1 -------------------------------------
\begin{figure*}[t]
\includegraphics[width=1.0\textwidth]{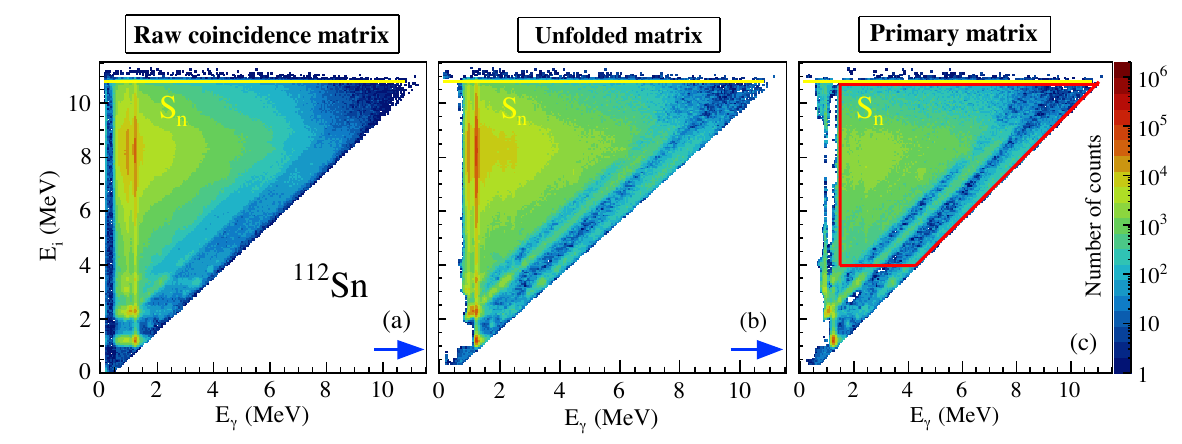}
\caption{\label{fig: matrices}  Experimental raw $p-\gamma$ coincidence (a), unfolded (b), and primary (c) matrices for $^{112}$Sn obtained in the ($p,p^{\prime}\gamma$) reaction. 
Yellow %dashed 
lines indicate the neutron separation energy of $^{112}$Sn.  
Red solid lines indicate the area of the primary matrix used in the Oslo method. The bin width is 124 keV for both axes. Blue arrows mark the sequence of the analysis steps.}
\end{figure*}
%--------------------------------------------------------------------------------

The energies and angles (relative to the beam direction) of emitted particles were recorded by the silicon particle telescope SiRi \cite{Guttormsen2011}, consisting of eight 1550-$\mu$m thick trapezoidal-shaped back $E$ detectors and 130-$\mu$m thick front $\Delta E$ detectors. 
Each front part is additionally segmented into eight strips with $\approx 2^{\circ}$ angular coverage, thus making up 64 $\Delta E$-$E$ combinations in total. 
SiRi was placed in a backward position with respect to the beam direction, covering angles from 126$^{\circ}$ to 140$^{\circ}$. 
This was primarily done to enhance the contribution from compound reactions relative to direct transfer reactions while also ensuring a larger transfer of angular momentum. 
Each SiRi detector had an $\approx 10.5$ $\mu$m Al foil in front to reduce the number of $\delta$ electrons. 
The energy resolution of the particle spectra depends primarily on the reaction channel, the beam-spot size, the target thickness, and the intrinsic energy resolution of SiRi. For the ($p,p^{\prime}$) channel, the full width at half maximum (FWHM) resolution was estimated from a Gaussian fit to elastically scattered protons to be $\approx  200$ keV, while using the first excited and ground states of $^{111}$Sn and $^{113}$Sn in the ($p,d$) and ($d,p$) channels yields the resolution of $\approx 320$ and $300$ keV, respectively.

To record $\gamma$ events, the target chamber was surrounded by the scintillator detector array CACTUS \cite{Guttormsen1990_CACTUS}, consisting of 28 spherically distributed 5$^{\prime\prime}\times$ 5$^{\prime\prime}$ NaI(Tl) scintillator $\gamma$-ray detectors. 
All of them were shielded with conical lead collimators to reduce the Compton contribution to the $\gamma$-ray spectra and to improve the peak-to-total ratio. 
The total efficiency of CACTUS  was measured with a $^{60}$Co source to be 15.2(1)\% ($E_{\gamma}$ = 1332 keV). 
The energy resolution of the NaI detectors at this $\gamma$-ray energy was $\approx 6.8$\%. 
The signals from the back detectors of SiRi were used as triggers for the data acquisition and the time of the NaI signals was recorded relative to the particle signals within a time window of  $\approx 1\mu$s. 

 The particle spectra were calibrated to known levels in the Sn isotopes populated in all three runs, whereas the spectra obtained for a 4 mg/cm2 thick natural Si target were used to calibrate $\gamma$ spectra. 
 The reaction channels of interest were further selected with the $\Delta E$-$E$ technique. 
 The kinematics of the studied reactions were used to convert particle energies within the selected channels into the corresponding excitation energies of $^{111,112,113}$Sn. 
 By gating on the prompt time peak and subtracting  background, we selected the desired particle-$\gamma$ events for the further analysis. 
 These events are presented in the form of a raw-data coincidence matrix shown in Fig.~\ref{fig: matrices}(a) for the case of $^{112}$Sn.

The $\gamma$-ray spectra were further corrected for the response functions of the CACTUS array \cite{Guttormsen1996}. 
The Compton subtraction method incorporated in the unfolding procedure allows for preserving the statistical fluctuations of the raw spectra in the resulting unfolded spectra without introducing any artificial features. 
Details of the procedure are outlined in Ref.~\cite{Guttormsen1996}. 
The unfolded matrix for $^{112}$Sn is shown in Fig.~\ref{fig: matrices}(b).

To extract the NLD and GSF from the coincidence data, the first-generation $\gamma$ rays from all possible cascades, i.e. stemming directly from each given initial excitation energy bin,  were singled out to form a so-called primary matrix (see Fig.~\ref{fig: matrices}(c)). 
This was done by means of the first-generation method described in detail in Ref.~\cite{Guttormsen1987}. This method exploits the assumption that $\gamma$ decay patterns of  excited levels are independent of the way of their formation, either through a direct population in a reaction or via decays of higher-lying excited states. It is expected to hold well for comparatively high excitation energy bins below the neutron threshold \cite{Guttormsen1987}. The distribution of primary $\gamma$ rays for each excitation energy bin is, thus, determined by subtracting a weighted sum of the spectra corresponding to the lower-lying excitation energy bins. This procedure has been  shown to be quite robust and provide reliable results \cite{Larsen11}. The primary matrix obtained in this way serves as the main input for the Oslo method. 

Prior to extracting the NLD and GSF from the first-generation spectra, we set a minimum limit $E_i^{min}$ for excitation energies  to ensure including the region of statistical decay only, while the upper limit is provided by the neutron separation energy $S_n$ (the outgoing neutrons were not measured). 
To exclude the low $\gamma$-ray energy regions affected by over- and under-subtraction of counts in the first-generation procedure, we also introduce an $E_{\gamma}^{min}$ limit. The regions used in this work for the further processing are given by  $3.0 \leq E_i \leq 8.2$ MeV, $E_{\gamma}\geq1.0$ MeV for $^{111}$Sn, $4.0 \leq E_i \leq 10.8$ MeV, $E_{\gamma}\geq1.5$ MeV for $^{112}$Sn, and  $5.5 \leq E_i \leq 7.7$ MeV, $E_{\gamma}\geq1.5$ MeV for $^{113}$Sn.
 
%--------------------------- Subsection 3: OSLO METHOD --------------------------
\section{\label{sec 3: Oslo method}Analysis with the Oslo method}

The core idea of the Oslo method lies in the decomposition of the primary matrix $P(E_{\gamma}, E_i)$ into the NLD $\rho_f=\rho(E_i-E_{\gamma})$ and the $\gamma$-transmission coefficient $\mathcal{T}_{i\rightarrow f}$:
\begin{equation}
\label{eq:1}
    P(E_{\gamma},E_i)\propto \rho_f\mathcal{T}_{i\rightarrow f}.
\end{equation}
This relation is based on a fact that the primary matrix is proportional to the probability of $\gamma$ decay of states within each initial excitation energy bin $E_i$ to the states of a final bin $E_f$ with $\gamma$-ray energies of $E_{\gamma}=E_i-E_f$. 
Both Fermi’s golden rule and the Hauser-Feshbach theory of statistical reactions can be used to provide the derivation of Eq.~\ref{eq:1} (Refs.~\cite{MIDTBO_1} and ~\cite{MIDTBO_2}, respectively). 
This decomposition holds in the same range of compound states as the first generation method. The dependence of the transmission coefficient on $E_i$, $E_f$, and $E_{\gamma}$ in Eq.~(\ref{eq:1}) significantly complicates factorization of two functions, $\rho_f$ and $\mathcal{T}$. To proceed with the decomposition, validity of the Brink-Axel hypothesis must be assumed \cite{Brink1955, Axel1962}. 
The generalized, most frequently used form of this hypothesis suggests the GSF to be solely a function of $\gamma$-ray energy, i.e. to be independent of spins, parities, and excitation energies of initial and final states. 
This effectively removes the excitation energy dependence of the $\gamma$-transmission coefficient $\mathcal{T}_{i\rightarrow f}\rightarrow \mathcal{T}(E_{\gamma})$. The applicability of this hypothesis has been previously discussed, e.g., in Refs.~\cite{Larsen11,Guttormsen16, Campo2018,Markova2021}.

\begin{figure}[t]
\includegraphics[width=1.0\columnwidth]{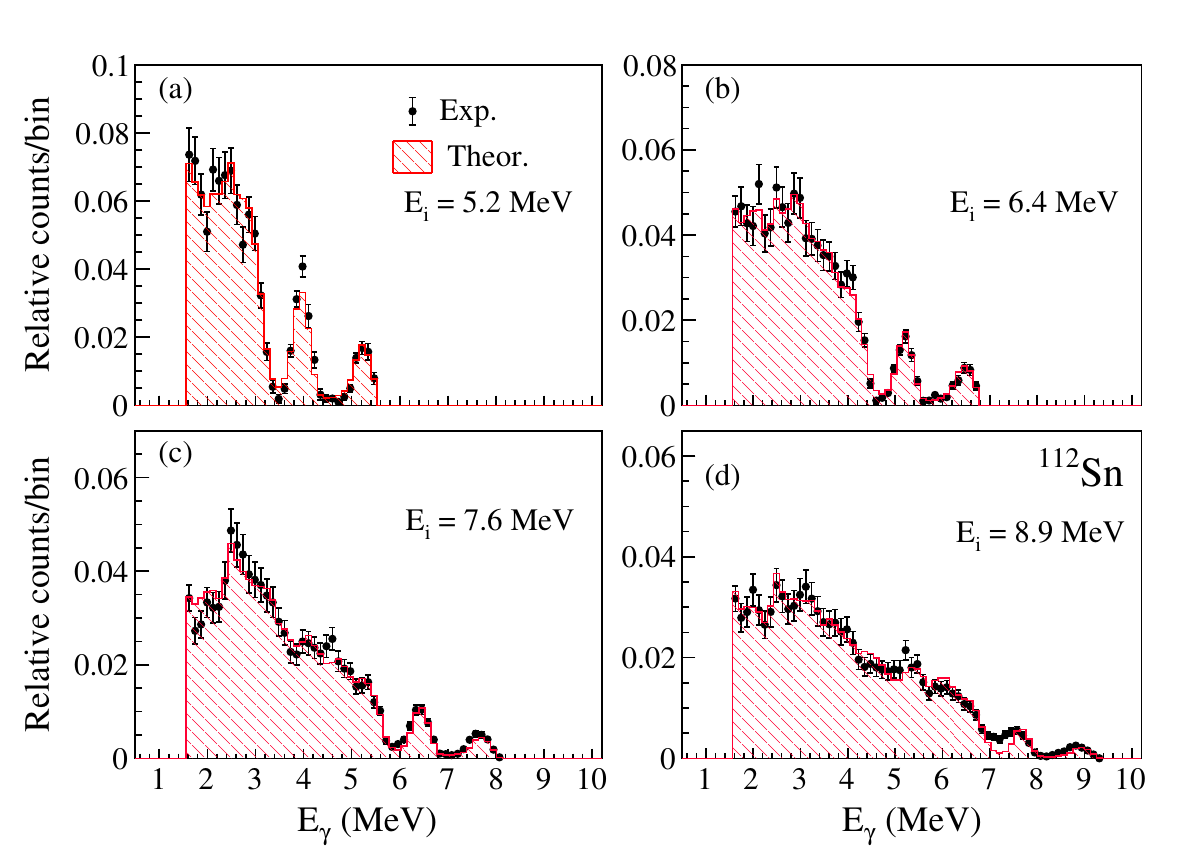}
\caption{\label{fig: P theor vs exp}
Experimental primary spectra for 5.2 MeV (a), 6.4 MeV (b), 7.6 MeV (c), and 8.9 MeV (d) excitation energy bins compared to the spectra predicted with the derived level density and $\gamma$-transmission coefficient [from Eq.~(\ref{eq:1})]. The excitation energy bins are 124-keV wide.
}
\end{figure}

The NLDs and the $\gamma$-transmission coefficients are obtained through an iterative $\chi^2$ procedure of fitting the experimental primary matrix (normalized to unity for each $E_i$) with a theoretical primary matrix given by
\begin{equation}
\label{eq:2}
P_{th}(E_{\gamma}, E_i)=\frac{\mathcal{T}(E_{\gamma})\rho(E_i-E_{\gamma})}{\sum_{E_{\gamma}=E_{\gamma}^{min}}^{E_i}\mathcal{T}(E_{\gamma})\rho(E_i-E_{\gamma})}.
\end{equation}
All the details of this procedure and the error propagation are described in Ref.~\cite{Schiller2000}. 
The obtained fit provides a very good agreement with the experimental primary matrix, as demonstrated for a few selected excitation energies in the case of $^{112}$Sn in Fig.~\ref{fig: P theor vs exp}. The theoretical function $P_{th}(E_{\gamma}, E_i)$ reproduces all experimental features quite well within a large interval of excitation energies below the neutron threshold. 

The fit given by Eq.~(\ref{eq:2}) yields the functional forms of the NLD and $\gamma$-transmission coefficient, i.e. their excitation energy and $\gamma$-ray energy dependencies, respectively. 
The general solutions $\tilde{\rho}(E_i-E_{\gamma})$ and $\tilde{\mathcal{T}}(E_\gamma)$ for both functions have the following forms \cite{Schiller2000}:
\begin{equation}
\label{eq:3}
\begin{split}
    \tilde{\rho}(E_i-E_{\gamma})=&A\rho(E_i-E_{\gamma})\exp[\alpha (E_i-E_{\gamma})],\\
    \tilde{\mathcal{T}}(E_\gamma)=&B\mathcal{T}(E_\gamma)\exp(\alpha E_{\gamma}),
\end{split}
\end{equation}
where $\rho(E_i-E_{\gamma})$ and $\mathcal{T}(E_\gamma)$ are two fixed solutions, $A$ and $B$ denote scaling coefficients, and $\alpha$ is a slope shared by $\rho(E_i-E_{\gamma})$ and $\mathcal{T}(E_\gamma)$. 
To determine the physical solutions of the NLDs and $\gamma$-transmission coefficients, one must apply external experimental information and model assumptions, as discussed in the following sections.

%------------------------------------ FIG_2 -------------------------------------
%--------------------------------------------------------------------------------

%----------------------- Subsection 3.1: NLD NORMALIZATION ----------------------

\subsection{\label{subsec 3.1: NLDs}Normalization of the level densities}

The key ingredients to determine the absolute value and the slope of the NLD  are the discrete low-lying levels and the value of the NLD at the neutron threshold, $\rho(S_n)$.  The most recent compilation of discrete levels \cite{ensdf} was used for all three isotopes. 
The neutron resonance spacings $D_0$ for $s$-wave neutrons or $D_1$ for $p$-wave neutrons from neutron resonance experiments are commonly used to estimate the $\rho(S_n)$ values. 
Among three isotopes studied in this work, only $^{112}$Sn is a stable nucleus with ground state spin and parity $J_t^{\pi_t}=0^+$ and can be used as a target in neutron resonance studies. For this reason, only $^{113}$Sn has readily available data on the resonance spacings \cite{Mughabghab18}.  
For the case of $s$-wave neutrons, levels of spin and parity $1/2^+$ of the residual $^{113}$Sn nucleus are populated, with the partial level density 
\begin{equation}
\label{eq:4}
    \frac{1}{D_0}=\frac{1}{2}\rho(S_n, J_t+1/2).
\end{equation}
Here, we utilize the procedure described in detail in Ref.~\cite{Larsen11}. 
Equal positive and negative parity contribution at $S_n$ is assumed, which is shown to be a reliable assumption at sufficiently high excitation energies (see e.g. \cite{Larsen11,Toft2010}). 
The spin and excitation energy dependence (denoted by $J$ and $E_x$, respectively) of the NLD are introduced through adopting the back-shifted Fermi gas (BSFG) form of the NLD from Ref.~\cite{Gilbert65}, $\rho(E_x,J)= \rho(E_x)g(E_x,J)$. 
The spin distribution $g(E_x, J)$ is expressed as a function of the energy-dependent spin-cutoff parameter $\sigma(E_x)$ \cite{Ericson58,Gilbert65}
\begin{equation}
\label{eq:5}
 g(E_x, J) \simeq \frac{2J+1}{2\sigma^2(E_x)}\exp\left[-\frac{(J+1/2)^2}{2\sigma^2(E_x)}\right].
\end{equation}
This allows for transforming Eq.~(\ref{eq:4}) into the relation for $\rho(S_n)$:
\begin{equation}
\label{eq:6}
\rho(S_n)=\frac{2\sigma^2}{D_0}\frac{1}{(J_t+1)\exp(-\frac{(J_t+1)^2}{2\sigma^2(E_x)})}.
\end{equation}
We chose the form of the spin-cutoff parameter at $S_n$ as given by Ref.~\cite{Gilbert65}:
\begin{equation}
\label{eq:7}
\sigma^2(S_n)=0.0888a\sqrt{\frac{S_n-E_1}{a}}A^{2/3},
\end{equation}
where $a$ and $E_1$ are the level-density and back-shift parameters for the BSFG model taken from global parametrizations of Ref.~\cite{Egidy05}. 
This choice of the spin-cutoff parameter is primarily motivated by observations made for the previously studied tin isotopes. Namely, the rigid-body form of the spin-cutoff parameter provides somewhat larger, overestimated values of $\rho(S_n)$ and, thus, the slopes of the experimental NLDs (relative to Eq.~(\ref{eq:7})).  Indeed, the effect of pairing correlations is expected to effectively reduce the moment of inertia as compared to the rigid-body model \cite{Hilaire2001}. The above-mentioned overestimation can be accounted for by using the Shape method \cite{Wiedeking2020} to constrain the true slope of the NLDs (see e.g. Ref.~\cite{Markova2022}). The limited experimental resolution of CACTUS for $^{111,113}$Sn and a too narrow range of useful Shape method data for $^{112}$Sn, however, prevent us from extracting reliable results with this method in these cases. Alternatively, a reduction factor can be applied to the rigid-body spin-cutoff parameter. To avoid introducing any additional parameters, we chose the form of $\sigma(S_n)$ given by Eq.~(\ref{eq:7}), corresponding to $\approx$ 80
\% of the rigid-body estimate. This choice is additionally supported by the previous analysis of $^{116,120,124}$Sn (see Ref.~\cite{Markova2021}), where the slopes of NLDs were obtained in a similar way and the respective slopes of GSFs were found to be in excellent agreement with the Coulomb excitation data \cite{Bassauer2020b}. This is also accounted for by an additional 10\% uncertainty we introduce for $\sigma(S_n)$ in this work.

%---------------------TABLE_1-------------------
\begin{table*}[t]
\caption{\label{tab:table_1}Parameters used for the normalization of the NLDs and GSFs for $^{111,112,113}$Sn.}
\begin{ruledtabular}
\begin{tabular}{lccccccccccccc}
Nucleus & $S_n$ & $D_0$ & $a$ & $E_1$ & $E_d$ & $\sigma_d$ & $\sigma(S_n)$ & $\rho(S_n)$ & $T$ & $E_0$ & $\langle\Gamma_{\gamma}\rangle$ \\ 
& (MeV) & (eV) & (MeV$^{-1}$) & (MeV) & (MeV) & & & ($ 10^5$ MeV $^{-1}$) &  (MeV) & (MeV) & (meV)\\
\noalign{\smallskip}\hline\noalign{\smallskip}
 $^{111}$Sn & 8.169 & 120(36)\footnotemark[1] & 12.05 & -0.29 & 1.08(7) & 2.7(4) & 4.6(5) & 3.5(13)\footnotemark[1] & 0.67$^{+0.03}_{-0.02}$ & -0.06$^{+0.04}_{-0.11}$  &  76(18)\footnotemark[1] \\
 
 $^{112}$Sn & 10.788  & 3(1)\footnotemark[1] & 12.53 & 1.12 & 2.83(4) & 2.8(4) & 4.8(5) & 24.6(8)\footnotemark[1] & 0.71$^{+0.02}_{-0.02}$ & 0.66$^{+0.09}_{-0.08}$ &  87(34)\footnotemark[2] \\
 
$^{113}$Sn & 7.744 & 172(10) & 12.77 & -0.27 & 1.88(2) & 3.5(7) & 4.6(5) & 2.5(5) & 0.63$^{+0.01}_{-0.01}$ & 0.20$^{+0.04}_{-0.04}$ &  73(8) \\
\end{tabular}
\end{ruledtabular}
\footnotetext[1]{From systematics.}
\footnotetext[2]{Modified (see text).}
\end{table*}
%-----------------------------------------------

Due to the lower limit of $\gamma$-ray energies mentioned in the previous section, the experimental NLDs do not reach the neutron threshold, but rather stop at energies $\approx 1-2$ MeV below $S_n$. To constrain the slope of the NLD, the experimental values have to be extrapolated to $\rho(S_n)$. 
Here, we use the constant-temperature model \cite{Ericson59,Gilbert65,Egidy05}:
\begin{equation}
\label{eq:8}
    \rho_{CT}(E_x)=\frac{1}{T_{{CT}}}\exp(\frac{E_x-E_0}{T_{{CT}}}),
\end{equation}
with the temperature ($T_{CT}$) and shift energy ($E_0$) treated as free parameters. 
This model was favored over the BSFG trend in the present cases based on the observed excitation energy dependencies and the quality of the $\chi^2$ fit to the experimental data. 
When the energy gap is relatively small ($\approx 1-2$ MeV), the choice of the extrapolation model is not expected to play any significant role as compared to other sources of uncertainties.

As mentioned previously, both $^{110}$Sn and $^{111}$Sn are unstable isotopes, and no experimental information on neutron resonance spacings is available for $^{111}$Sn and $^{112}$Sn.
Hence, the values of the NLD at $S_n$ were obtained from the systematics available for stable Sn isotopes in the same way as described in Ref.~\cite{Markova2022} with the spin-cutoff parameter given by Eq.~(\ref{eq:7}). We additionally include the above-mentioned  10\% error for $\sigma(E_x)$ in the total errors of $\rho(S_n)$ for each isotope in the systematics together with the experimental uncertainties of $D_0$. 

The obtained error bands of the  NLDs include  statistical errors combined with the systematic errors from the unfolding and the first-generation method, and are calculated according to the procedure from Ref.~\cite{Schiller2000}. 
For $^{113}$Sn, the 10\% error of the spin-cutoff parameter is propagated together with the experimental error of $D_0$ into the NLD uncertainty at the neutron separation energy and also included in the systematic error band as was done previously in Ref.~\cite{Kullmann2019}. We note that if we would use the predictions from systematics for $^{113}$Sn rather than the neutron resonance data, the normalization parameters would be slightly lower but well within the error bars reported here. The $D_0$ values for $^{111,112}$Sn were estimated from the $\rho(S_n)$ values extracted from the systematics. We assume a 30\% error of $D_0$ in both cases, which is approximately twice as large as the largest experimental error of $D_0$ available for other Sn isotopes. A good agreement, well within the estimated error bands, between the slopes of the obtained GSFs with the ($p,p^{\prime}$) Coulomb excitation data (see Sec.~\ref{sec 5: GSF}) also supports this choice. These errors were combined with the $\sigma(E_x)$ uncertainties and propagated in the total systematic error bands for the NLDs of $^{111,112}$Sn. All parameters for the NLD normalization used in this work are presented in Table \ref{tab:table_1}. 

%----------------------- Subsection 3.2: GSF NORMALIZATION ----------------------

\subsection{\label{subsec 3.2: GSFs}Normalization of the $\gamma$-ray strength functions}

The slope of the $\gamma$-transmission coefficient, also defined by the parameter $\alpha$ (see previous sections), is automatically determined through normalizing the NLD. 
The only parameter left to be constrained is $B$, i.e. the absolute value of $\mathcal{T}(E_\gamma)$.
To extract this parameter, we utilize the expression for the average radiative width $\langle\Gamma(E_x, J, \pi)\rangle$ for the levels of spin-parity $J^{\pi}$ at excitation energy $E_x$ \cite{Kopecky1990}:
\begin{equation}
\label{eq:9}
\begin{split}
    \langle\Gamma(E_x, J, \pi)\rangle=&\frac{1}{2\pi\rho(E_x, J, \pi)}\sum_{XL}\sum_{J_f,\pi_f}\int_{E_{\gamma}=0}^{E_x}dE_{\gamma}\times\\&\times\mathcal{T}_{XL}(E_\gamma)\rho(E_x-E_{\gamma},J,\pi),
\end{split}
\end{equation}
with $X$ and $L$ being the electromagnetic character and multipolarity of the $\gamma$ radiation. The latter can be safely assumed to be of dipole nature in our case ($E1+M1$, see e.g. \cite{Kopecky1990}). 
The GSF, $f(E_{\gamma})$, is then directly obtained from the $\gamma$-transmission coefficient by the relation  $B\mathcal{T}(E_{\gamma})= 2\pi E^3f(E_{\gamma})$ \cite{Belgya2006}. 

The total average radiative width $\langle\Gamma_{\gamma}\rangle$ obtained from $s$-wave neutron capture experiments [corresponds to $\langle\Gamma(S_n, J_t, \pi_t)\rangle$ in Eq.~(\ref{eq:9})] can be used to find the scaling parameter $B$. 
We adopt the prescription of Ref.~\cite{Capote2009} and use the following excitation energy dependence of the spin-cutoff parameter:
\begin{equation}
\label{eq:10}  
    \sigma^2(E_x) = \sigma_d^2 + \frac{E_x-E_d}{S_n-E_d}[\sigma^2(S_n)-\sigma_d^2],
\end{equation}
with $\sigma_d$ estimated from the discrete lower-lying levels at $E_x\approx E_d$~\cite{ensdf}.

For $^{113}$Sn, the $\langle\Gamma_{\gamma}\rangle$ value at $S_n$ is available from  $s$-wave neutron resonance studies \cite{Mughabghab18}. 
For $^{111}$Sn and $^{112}$Sn, however, these values have to be constrained from the systematics for other Sn isotopes as it was done for $^{124}$Sn in Ref.~\cite{Markova2022}. 
The value of $\langle\Gamma_{\gamma}\rangle = 76(18)$ meV obtained in this way for $^{111}$Sn seems to be quite satisfactory based on the comparison with the ($p,p^{\prime}$) Coulomb excitation data, while the $\langle\Gamma_{\gamma}\rangle = 121(22)$ meV value for $^{112}$Sn yields a significantly overestimated GSF. 
Given the good agreement of the Oslo data with the ($p,p^{\prime}$) Coulomb excitation strengths for other even-even Sn isotopes (see \cite{Markova2021}), we chose to apply an additional reduction factor to the $\langle\Gamma_{\gamma}\rangle$ value for $^{112}$Sn extracted from the systematics. 
This factor is obtained through a $\chi^2$-minimization with our GSF and the ($p,p^{\prime}$) data below the neutron threshold. 
The  $\langle\Gamma_{\gamma}\rangle$ from the systematics is set to be the maximum value, spanning a symmetrical error bar for $\langle\Gamma_{\gamma}\rangle$ in $^{112}$Sn. For the $^{111}$Sn nucleus this error is provided by the fit error from the systematics. 

The error bands shown for the GSFs in Sec.~\ref{sec 5: GSF} comprise the statistical errors, systematic errors of the unfolding and the first-generation procedure, as well as the propagated errors due to the $D_0$, $\langle\Gamma_{\gamma}\rangle$, $\sigma(S_n)$, $\sigma_d$, and $E_d$ values.
All parameters and their uncertainties used in the normalization of the GSFs are listed in Table~\ref{tab:table_1}. 

%------------------------------- Section 3.1: NLDs ------------------------------

\section{\label{sec 4: NLD+Thermo}Nuclear level densities and thermal properties}

\begin{figure}[t]
\includegraphics[width=1.0\columnwidth]{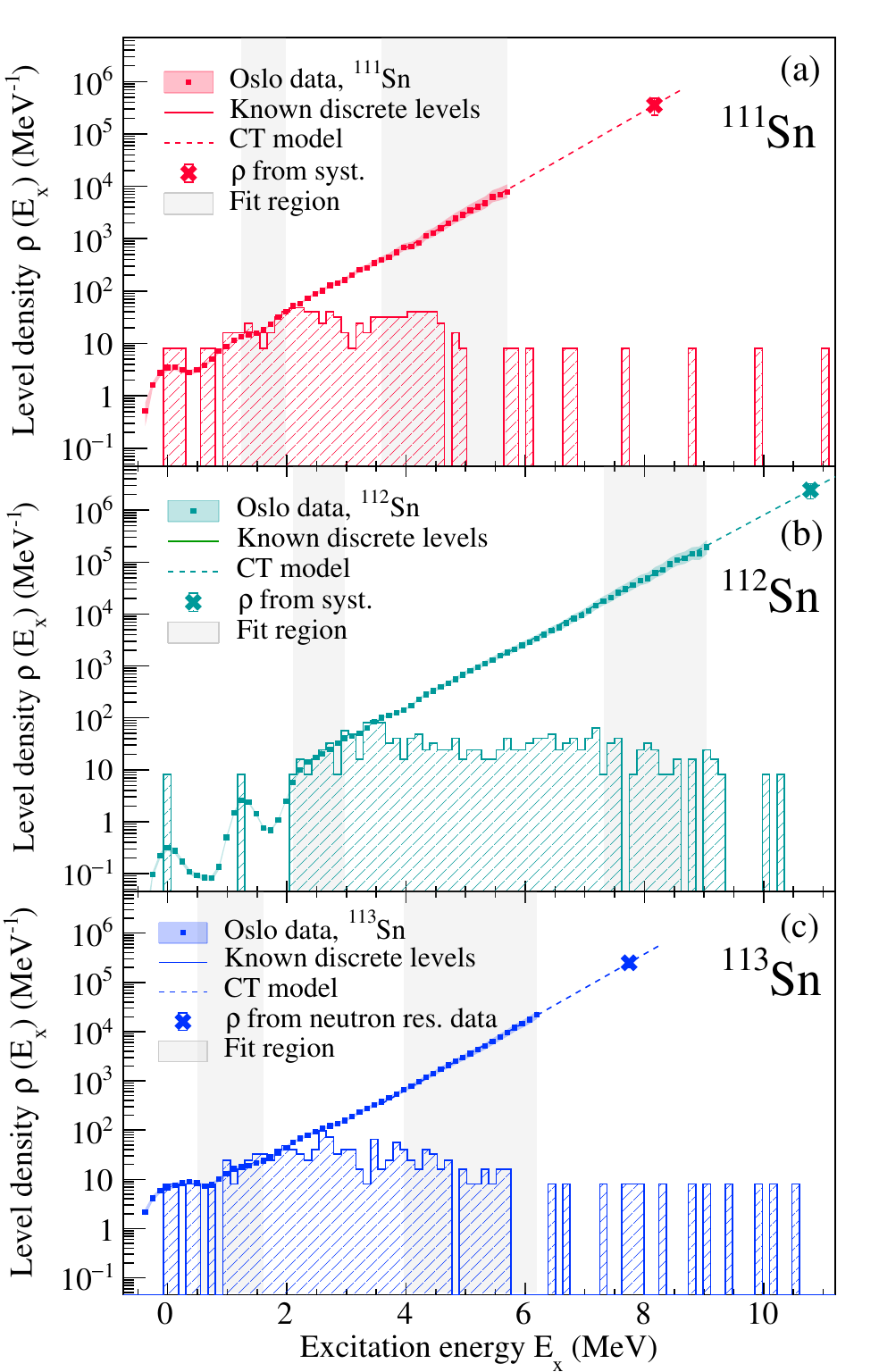}
\caption{\label{fig: LDs}
Experimental NLDs of $^{111}$Sn (a), $^{112}$Sn (b), and $^{113}$Sn (c). The $\rho(S_n)$ values are marked as crosses, discrete levels are presented as hatched histograms. The gray shaded areas mark the lower and higher excitation energy normalization regions.
}
\end{figure}

The NLDs of $^{111,112,113}$Sn extracted with the Oslo method are presented in Fig.~\ref{fig: LDs}. All NLDs follow nicely a number of low-lying excited states up to $\approx$2.2 MeV for $^{111}$Sn, 3.5 MeV for $^{112}$Sn, and 2.7 MeV for $^{113}$Sn. Up to these energies the level schemes can be, thus, considered complete. 
As compared to $^{113}$Sn, the result for $^{111}$Sn slightly underestimates the experimental NLD below $\approx$ 1 MeV, most likely, due to the difference in the reaction mechanism and energy, favouring  higher momentum transfer in the ($p,d$) reaction \cite{Blankert1981}. 
With the typical resolution of $\approx$ 200-300 keV, only the ground state and the first excited state of $^{112}$Sn are clearly separated. 
The non-zero values of the NLD between these states are due to the above-mentioned experimental resolution and some leftover counts remaining between the diagonals of the primary matrix after the background subtraction and unfolding. 
The lowest-lying levels in odd-even isotopes are seen as a single bump below $E_x =500$ keV in $^{111}$Sn and 700 keV in $^{113}$Sn.  
At energies above $E_x\approx$ 4 MeV, all nuclei demonstrate a steep exponential increase toward the neutron threshold, following a constant-temperature trend. 
For this reason, the normalization fit needed to constrain the CT model parameters for the extrapolation of the NLDs was found to be quite insensitive to the exact choice of the normalization limits (marked as shaded gray areas in Fig.~\ref{fig: LDs}).

%---------------------------------------- FIG_3 ----------------------------------------
%---------------------------------------------------------------------------------------

\begin{figure}[t]
\includegraphics[width=1.0\columnwidth]{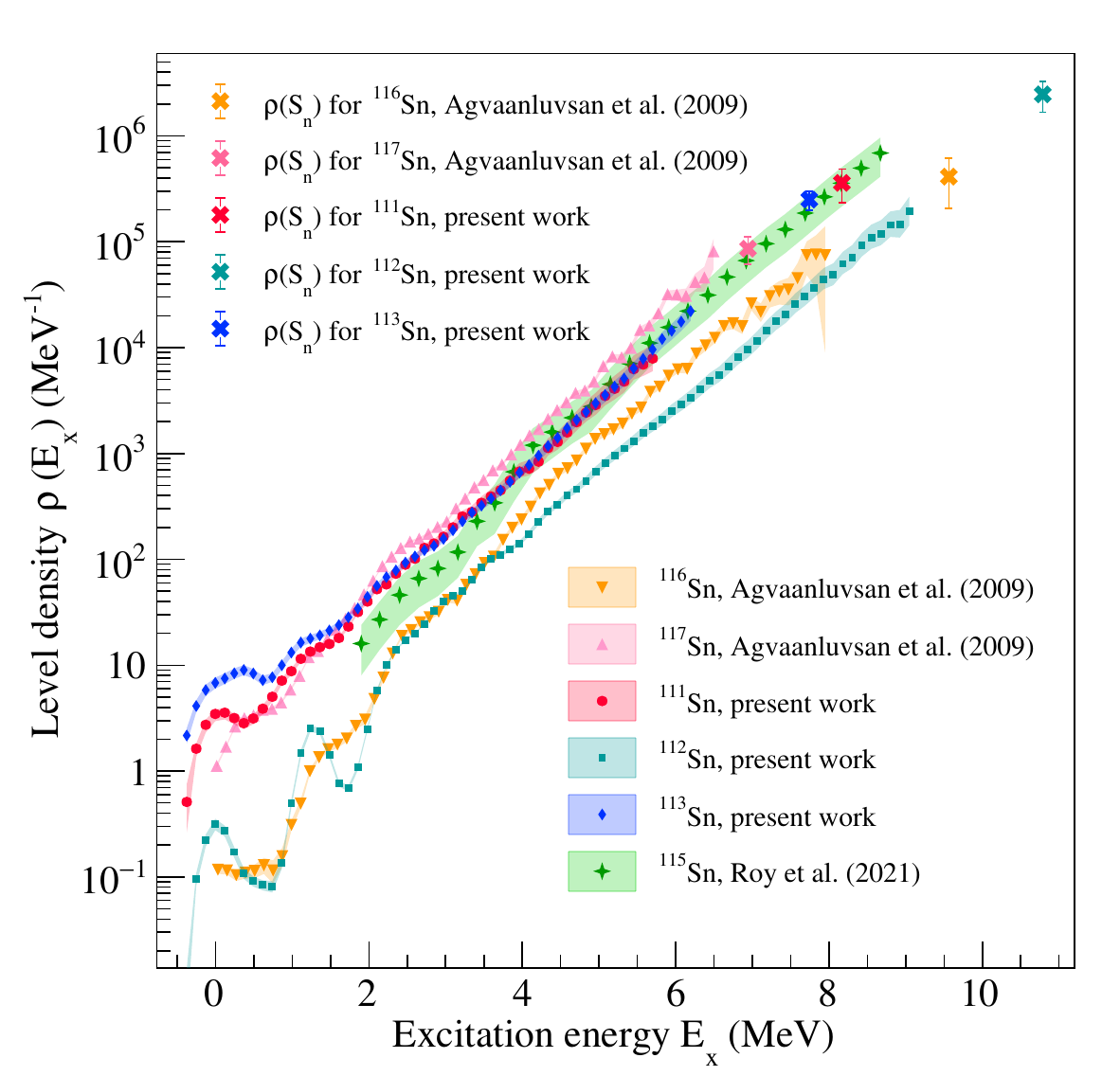}
\caption{\label{fig: Sn LDs}
Comparison of the experimental nuclear level densities for $^{115}$Sn \cite{Roy2021}, $^{116}$Sn \cite{Agvaanluvsan09-1}, $^{117}$Sn \cite{Agvaanluvsan09-1}, shown together with the $\rho(S_n)$ values, and the present data for $^{111,112,113}$Sn.
}
\end{figure}

The comparison of the experimental results for $^{111,112,113}$Sn with other neighboring Sn isotopes is shown in Fig.~\ref{fig: Sn LDs}.  Here, we include the NLDs of $^{115}$Sn, studied in a neutron evaporation experiment \cite{Roy2021}, and $^{116,117}$Sn, studied with the Oslo method in ($^3$He, $^3$He $\gamma$) and ($^3$He, $^4$He $\gamma$) experiments \cite{Agvaanluvsan09-1}.  The level densities of all shown odd-even isotopes are by a factor of 5-9 higher than those of the even-even isotopes, as expected due to the presence of an uncoupled valence neutron in the odd-even nuclei. 
The NLDs of $^{111}$Sn and $^{113}$Sn agree quite well within the estimated error bands with each other above $\approx 2$ MeV. Moreover, their slopes and absolute values agree above $\approx$ 3.5 MeV with those of the NLD in $^{115}$Sn~\cite{Roy2021}. Similarly, the corresponding $\rho(S_n)$ estimates lie well within the error band of the neutron evaporation experiment. 
The same is true for the $\rho(S_n)$ value of $^{117}$Sn, which, however, appears to be higher in absolute values below the neutron threshold than all other odd-even isotopes. As no considerable structural changes in these odd-even isotopes are predicted, we do not expect any significant change in the observed slopes. 
The NLD of $^{117}$Sn being slightly higher might be indeed due some minor systematic evolution of the NLD with an increasing neutron number. 
However, it is important to mention that the BSFG model was used for the extrapolation in the case of $^{117}$Sn. It usually tends to slightly increase the NLD values when approaching the neutron threshold (see Ref.~\cite{Agvaanluvsan09-1} and Fig.~8 in Ref.~\cite{Guttormsen2015}). In many studied cases, including $^{111,112,113}$Sn, the BSFG yields a poorer $\chi^2$ score for the fit at high excitation energies, while the CT model provides a rather good fit in the same energy range and reproduces the NLD quite well below these energies. For example, within an energy range between 8 and 9 MeV in $^{112}$Sn, the fit provided by the BSFG results in a $\chi^2$ score which is a factor of six worse than the one obtained with the CT model.

%---------------------------------------- FIG_4 ----------------------------------------
%---------------------------------------------------------------------------------------
%---------------------------------------- FIG_5 ----------------------------------------
\begin{figure}[t]
\includegraphics[width=1.0\columnwidth]{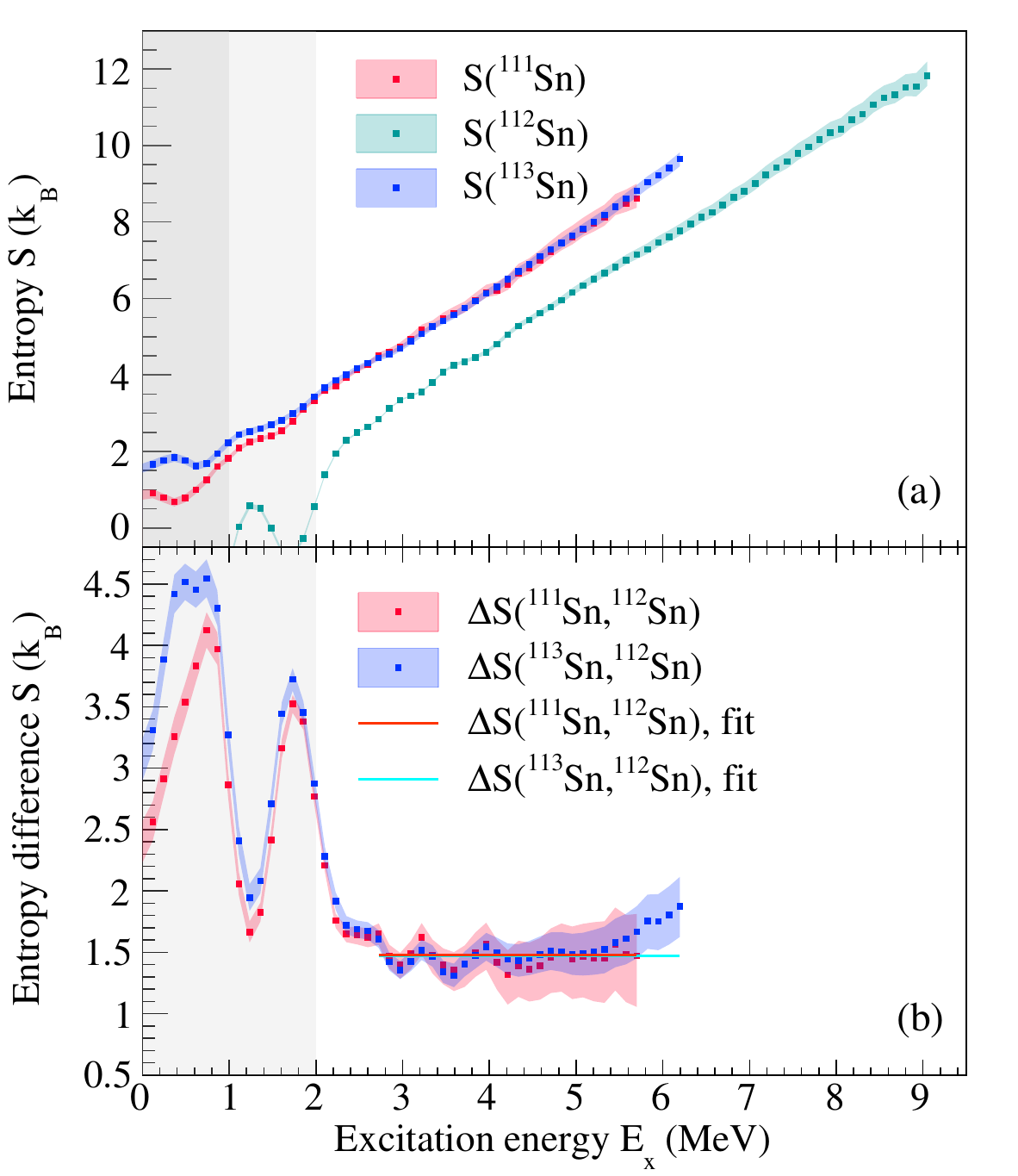}
\caption{\label{fig: Entropies}
Experimental entropies for $^{111,112,113}$Sn (a) and entropy differences $\Delta S(^{111}$Sn --$^{112}$Sn) and $\Delta S(^{113}$Sn --$^{112}$Sn) (b). Light and darker gray-shaded areas below 2 and 1 MeV indicate the areas where the entropies for $^{112}$Sn and $^{111,113}$Sn, respectively, are disregarded. Horizontal lines correspond to $\chi^2$ fits with constant functions.
}
\end{figure}
%---------------------------------------------------------------------------------------
The different approach for the extrapolation of the NLD to $\rho(S_n)$ might also be the main explanation for the difference in absolute values of the NLDs in $^{116}$Sn and $^{112}$Sn. This is additionally supported by the $\rho(S_n)$ value from Ref.~\cite{Agvaanluvsan09-1}, which seems to agree well with the CT slope predicted for $^{112}$Sn. 
Otherwise, the two NLDs follow the same trend below $\approx$ 3.5 MeV. 
As an older version of the particle telescope with a worse energy resolution was used in the earlier experiments (see Ref.~\cite{Agvaanluvsan09-1}), the ground and the first excited state at 1.293 MeV of $^{116}$Sn are rather seen as two consecutive bumps around the ground state and the 1.256 MeV-state of $^{112}$Sn.
Overall, the NLDs of $^{111,112,113}$Sn are considerably smoother and more featureless than those of $^{116,117}$Sn at relatively high excitation energies, which is most likely due to the better statistics of the newer experiments. 

To study possible structural features present in the NLDs, we extract the entropies $S$ and temperatures $T$, similar to how it was done in Refs.~\cite{Agvaanluvsan09-1, Nyhus2012, Roy2021}. Here, we have chosen to assume that these nuclei, given the experimental conditions, can be described within the microcanonical approach. By definition, the microcanonical entropy $S(E_x)$ is defined by the number of different ways a system can be arranged and, thus, can be derived through the corresponding partition function, namely the multiplicity of the populated states $\Omega_s$:
\begin{equation}
\label{eq:11}  
    S(E_x)=k_B\ln{\Omega_s(E_x)},
\end{equation}
where $k_B$ is the Boltzmann constant. 
To link this to the experimental NLD, one has to have an access to  the distribution of the populated spins at each excitation energy or, as in Ref.~\cite{Agvaanluvsan09-1,Nyhus2012}, introduce an averaged factor so that:
\begin{equation}
\label{eq:12}  
    \Omega_s(E_x) = (2\langle J(E_x) \rangle +1)\Omega_l,
\end{equation}
where $\langle J \rangle$ is the average populated spin and $\Omega_l$ is the multiplicity of levels. 
As the exact spin distribution is quite uncertain and since we are interested in the excitation energy dependence of $S(E_x)$ rather than its absolute values, we omit the spin-dependent factor.
Additionally, we introduce a parameter $\rho_0$ so that
\begin{equation}
\label{eq:13}  
    \Omega_l(E_x) = \frac{\rho(E_x)}{\rho_0}
\end{equation}
and the entropy of the even-even $^{112}$Sn at the ground state equates to zero, as expected. In the earlier works, the $\rho_0$ value was chosen such that the entropy at the excitation energy bin around 0 MeV yields $S\sim 0$ $k_B$ \cite{Agvaanluvsan09-1,Nyhus2012}. 
As the experimental NLD underestimates the theoretical one at the ground state, we chose $\rho_0$ to be the average of the ground state and the first $2^+$ state densities, 
%(the tabulated NLDs are equal), 
$\rho_0=1.431$ 1/MeV. 
The same value was taken for the odd-even isotopes as well. This choice does not affect the main trends of interest in the excitation energy dependence of $S(E_x)$ \cite{Agvaanluvsan09-1,Nyhus2012,Roy2021}.

The experimental entropies of $^{111,112,113}$Sn are shown in Fig.~\ref{fig: Entropies} (a). As $S(E_x)$ is not defined for excitation energy bins with no discrete states, we disregard the excitation energy ranges below 1 MeV for $^{111,113}$Sn (dark-gray area) and below 2 MeV for $^{112}$Sn (light-gray area), including all such bins. Following Ref.~\cite{Guttormsen2001}, we apply an assumption that the change in entropy between systems with an unpaired valence neutron ($^{111,113}$Sn) and a system with only paired neutrons ($^{112}$Sn) can be described with a constant shift as a function of excitation energy. The entropy differences $\Delta S = S(^{111,113}$Sn)$-S(^{112}$Sn) above 2 MeV are shown in the lower panel of Fig.~\ref{fig: Entropies}. 
Both differences are quite similar, except for a slight increase closer to 6 MeV in case of $^{113}$Sn. 
Between $\approx 2.7$ and $5.5-6.0$ MeV both differences can be considered almost constant within the estimated error bands, with the average values of $\Delta S = 1.48^{+0.04}_{-0.02}$~$k_B$ and $1.47^{+0.02}_{-0.02}$~$k_B$ for $^{111}$Sn and $^{113}$Sn, respectively.
These values are slightly lower than those obtained for $^{116,117}$Sn ($\approx 1.6$ $k_B$) \cite{Agvaanluvsan09-1} and $^{118,119}$Sn (1.7(2) $k_B$) \cite{Toft2010}. 
The estimate presented in the latter study should be treated with care due to quite poor statistics of the experiments. 
Both works compare the $\Delta S$ values with the semi-empirical study on entropies of midshell nuclei in the rare-earth region, providing an averaged value of $\Delta S\approx1.7$ $k_B$~\cite{Guttormsen2001}. 
This study, however, does not include such light isotopes of Sn as $^{111,112}$Sn. 
Overall, the entropy differences of nuclei in the vicinity of $Z=50$ demonstrate quite a large spread in values, from $\approx 1$ to 2 $k_B$, and, therefore, the estimates we obtained for $^{111,112,113}$Sn are in accordance with this study as well as the earlier works on $^{116-119}$Sn.

%---------------------------------------- FIG_6 ----------------------------------------
\begin{figure}[t]
\includegraphics[width=1.0\columnwidth]{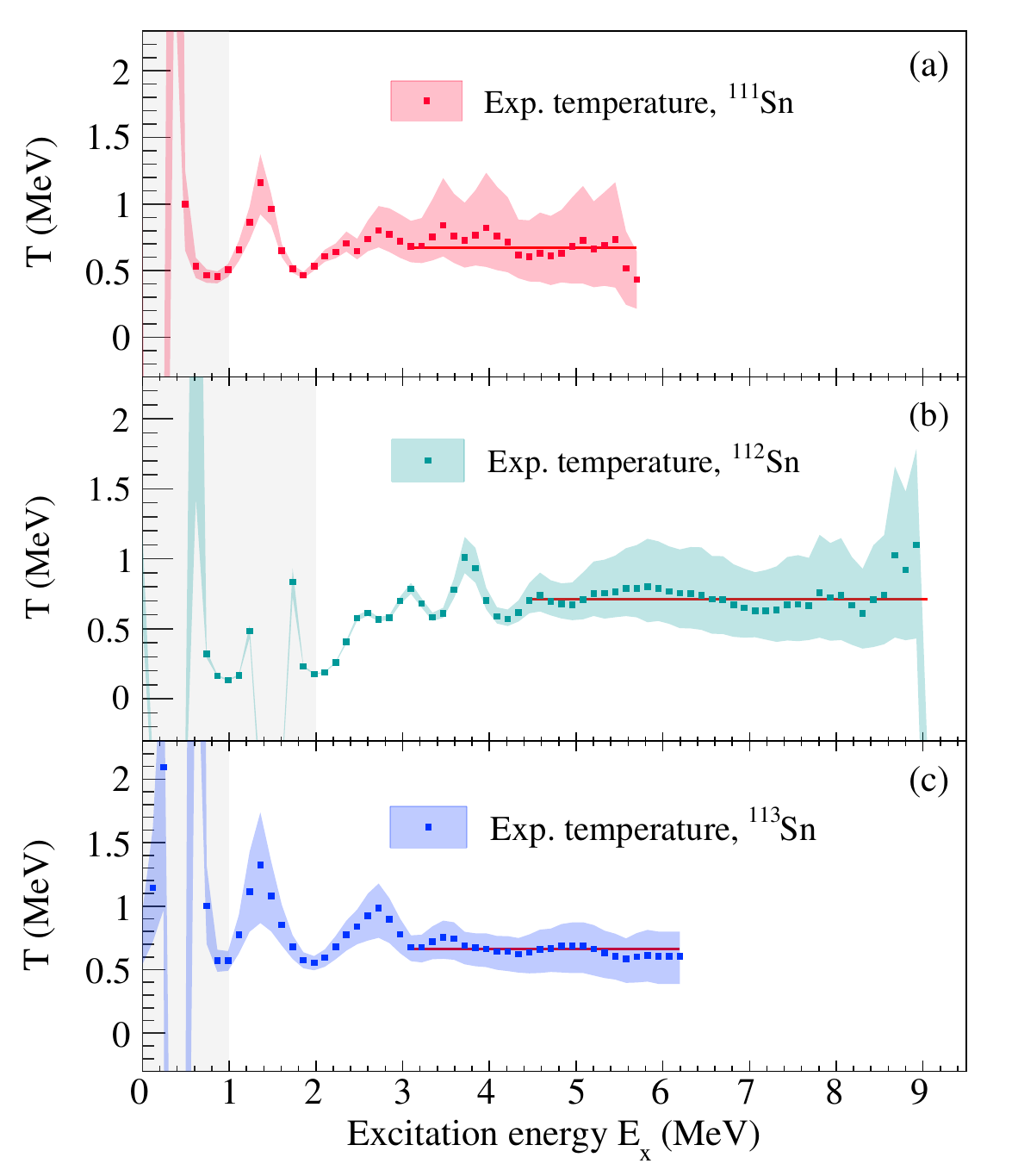}
\caption{\label{fig: Temperatures}
Experimental temperatures for $^{111}$Sn (a), $^{112}$Sn (b), and $^{113}$Sn (c). The gray-shaded areas below 2 in $^{112}$Sn and below 1 MeV in $^{111,113}$Sn indicate the areas where temperatures are not defined. Red solid lines denote the constant-temperature fits in each case.
}
\end{figure}
%---------------------------------------------------------------------------------------

The experimental entropies might potentially be used to shed some light on the process of Cooper pair breaking, contributing to the formation of levels in the NLD. According to the microscopic calculations with seniority-conserving and non-conserving interactions, this process is seen as step-like structures of the NLDs, experimentally observed for $^{56,57}$Fe and $^{96,97}$Mo \cite{Schiller2003}. Similar structures are clearly seen in $^{116-119}$Sn \cite{Agvaanluvsan09-1,Toft2010}, where the features  at relatively high excitation energies should be considered with care due to very large experimental error bars. A few quite clear features are seen in the entropy of $^{115}$Sn at $\approx 2-3$ and $4-5$ MeV \cite{Roy2021}. 

To amplify and study all subtle variations of the entropy, it is convenient to extract the microcanonical temperature $T(E_x)$:
\begin{equation}
\label{eq:14}  
T(E_x) = \left(\frac{\partial S(E_x)}{\partial E_x} \right)^{-1}.
\end{equation}
The resulting temperatures for all three isotopes are displayed in Fig.~\ref{fig: Temperatures}. 
The gray-shaded areas below $1$ MeV in $^{111,113}$Sn and 2 MeV in $^{112}$Sn correspond to the disregarded ranges of entropies, similarly to those presented in Fig.~\ref{fig: Entropies}. For all cases, the first bumps at $\approx 1.0-1.8$ MeV in the temperatures reflect the change of the NLD and entropy slope with the onset of a large amount of states above 1 MeV. 
These states are expected to be of predominantly single-particle (also featuring an uncoupled neutron) and collective nature. 
A similar effect of the collective states can be seen as a bump between 2 and 3 MeV in $^{112}$Sn. 
The next clear feature in $^{113}$Sn is at $\approx 2.6-3.0$ MeV. 
This energy is quite close to the double neutron pair-gap energy of $\approx 2.6$ MeV, and this bump can be a candidate for the first broken neutron Cooper pair. 
An analogous peak in the same energy range of $^{111}$Sn is somewhat less prominent, primarily due to the poorer statistics of the ($p,d \gamma$) experiment. 
Similarly, in $^{115}$Sn this feature is quite clear in the temperature profile \cite{Roy2021}. 
However, the peak at $4-5$ MeV in $^{115}$Sn is seen neither in $^{111}$Sn nor in $^{113}$Sn. 
Instead, these nuclei demonstrate almost constant-temperature regimes already above $\approx 3$ MeV with the average temperature of $T=0.67^{+0.06}_{-0.04}$ MeV for $^{111}$Sn and $T=0.66^{+0.03}_{-0.03}$ MeV for $^{113}$Sn, well in agreement with the corresponding temperatures from the CT extrapolation in Table \ref{tab:table_1}. 
This might be partly due to the experimental resolution, which smears subtle features of the NLD, no longer visible in the temperature profile. 
In addition, the process of Cooper pair breaking becomes more continuous at higher excitation energies, resulting in the constant-temperature behavior. 
As the proton $Z=50$ shell is closed, the contribution of breaking proton pairs is expected to begin at higher energies, above $\approx 4$ MeV. 

For the case of $^{112}$Sn, the temperature profile is quite similar to the odd nuclei. The most prominent feature is a peak at $3.6-4.0$ MeV which might again correspond to the first broken neutron pair ($2\Delta_n \approx 3.0$ MeV), given an extra energy needed to form a new configuration with the unpaired neutrons. The constant-temperature regime sets in above $\approx 4.5$ MeV with the average value $T=0.71^{+0.04}_{-0.03}$, which is in accordance with the fit temperature from Table \ref{tab:table_1}. 
In general, all of the above-mentioned trends are quite consistent with the previously published works on $^{115-119}$Sn, supporting the interpretation of the most prominent features of the NLDs in $^{111-113}$Sn.

%------------------------------- Section 3.1: GSFs ------------------------------

\section{\label{sec 5: GSF}$\gamma$-ray strength functions}

\begin{figure}[b]
\includegraphics[width=1.0\columnwidth]{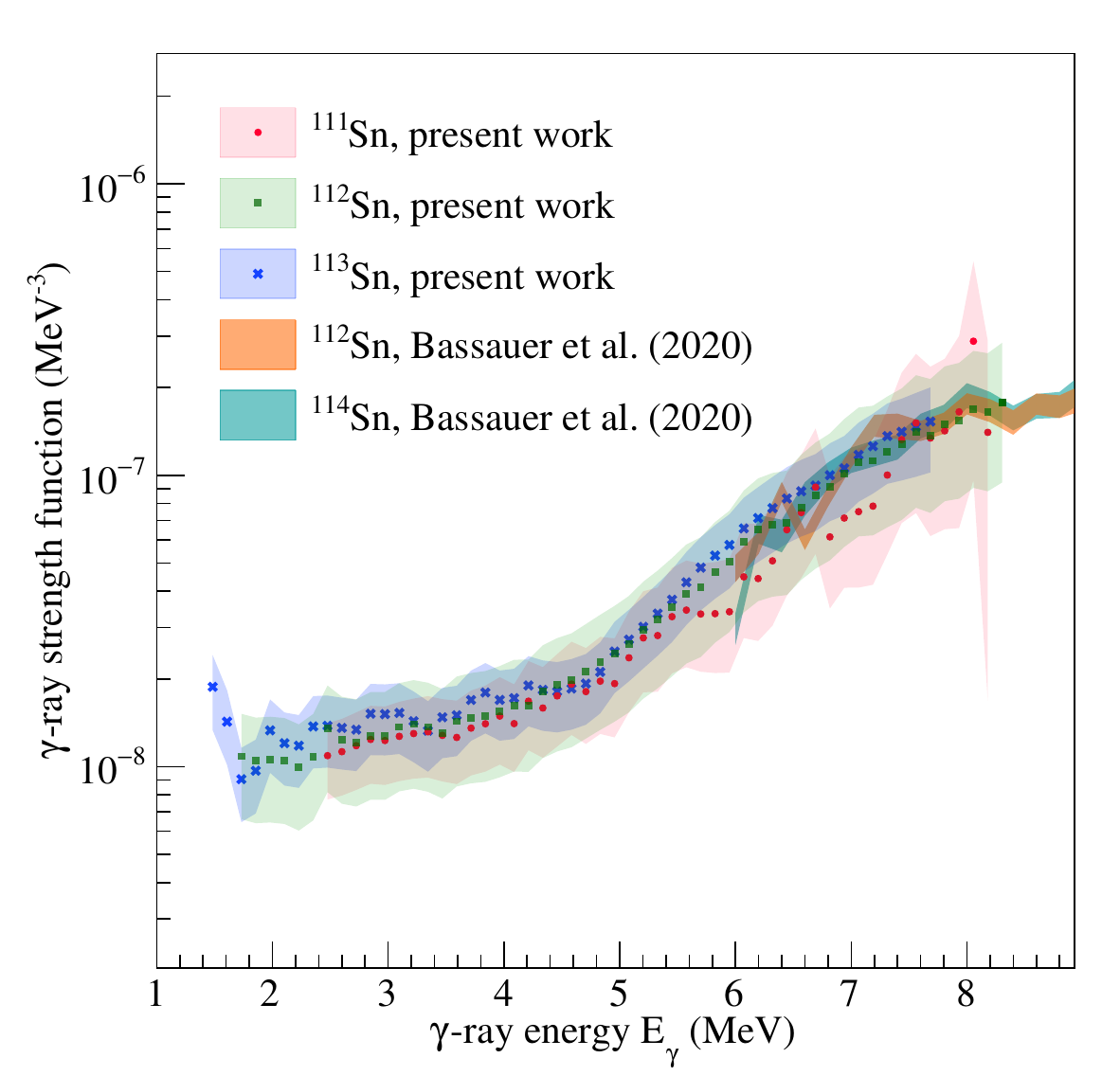}
\caption{\label{fig: Sn GSFs}
The experimental GSFs for $^{111,112,113}$Sn, shown together with the ($p,p^{\prime}$) Coulomb excitation data for $^{112,114}$Sn \cite{Bassauer2020b}.
}
\end{figure}

The experimental dipole GSFs of $^{111,112,113}$Sn extracted with the Oslo method are displayed in Fig.~\ref{fig: Sn GSFs}. 
The GSF of $^{112}$Sn above 8.3 MeV is not shown due to very poor statistics in the primary matrix at high $\gamma-$ray energies. 
All strengths agree well within the estimated error bands not only in slopes, but also in absolute values, demonstrating similar trends for the shown energy range. 
Even though the data points of $^{111}$Sn suffer from relatively low statistics above $\approx 6$ MeV, they still remain in good agreement with the GSFs of $^{112,113}$Sn up to the neutron threshold. 
This behavior is expected due to the similar structural properties of the studied isotopes.

The earlier published cases \cite{Agvaanluvsan2009-2, Toft2010, toft2011} have been primarily compared to each other and various $(\gamma, n)$ data above the neutron separation energy. The recent series of Coulomb excitation experiments through the ($p,p^{\prime}$) reaction performed on even-even $^{112,114,116,118,120,124}$Sn \cite{Bassauer2020b} provide us with GSFs below and above the neutron threshold and, thus, an excellent opportunity to compare and benchmark the slopes and absolute values of our GSFs below $S_n$. Figure \ref{fig: Sn GSFs} displays the comparison of our results with the GSFs extracted from the ($p,p^{\prime}$) spectra for $^{112,114}$Sn. 
The peak-like feature at $\approx 6.4$ MeV in the ($p,p^{\prime}$) data is not  seen in the Oslo strengths, likely due to the significantly worse experimental resolution.  

%---------------------------------------- FIG_7 ----------------------------------------
%---------------------------------------------------------------------------------------

An excellent agreement within the error bars of the Oslo-method GSF for $^{113}$Sn with those for $^{112,114}$Sn above 6 MeV supports the assumptions made to normalize this strength. A similar agreement for  $^{111}$Sn further supports the approach to assess the missing normalization parameters $\rho(S_n)$ and $\langle\Gamma_{\gamma}\rangle$ from the systematics. As previously mentioned, the latter parameter for $^{112}$Sn was estimated by scaling the Oslo method GSF to the ($p,p^{\prime})$ data. A good agreement of all strengths in slopes suggests that such scaling is needed due to, most probably, the systematics failing to reproduce a reasonable value of $\langle\Gamma_{\gamma}\rangle$ for $^{112}$Sn. 

Due to the overlap with the ($p,p^{\prime}$) data, covering also the energy range above the neutron threshold including the isovector giant dipole resonance (IVGDR), we are able to quantify the low-lying $E1$ strength in $^{111,112,113}$Sn, similarly to how it was done in Refs.~\cite{Agvaanluvsan2009-2,toft2011}. 
In the earlier publications this strength was referred to as the pygmy dipole resonance (PDR). However, due to the lack of experimental information on the isovector or/and isoscalar nature of this strength in the present cases we prefer to use a more general term of a low-lying $E1$ strength.

\begin{figure}[t]
\includegraphics[width=1.0\columnwidth]{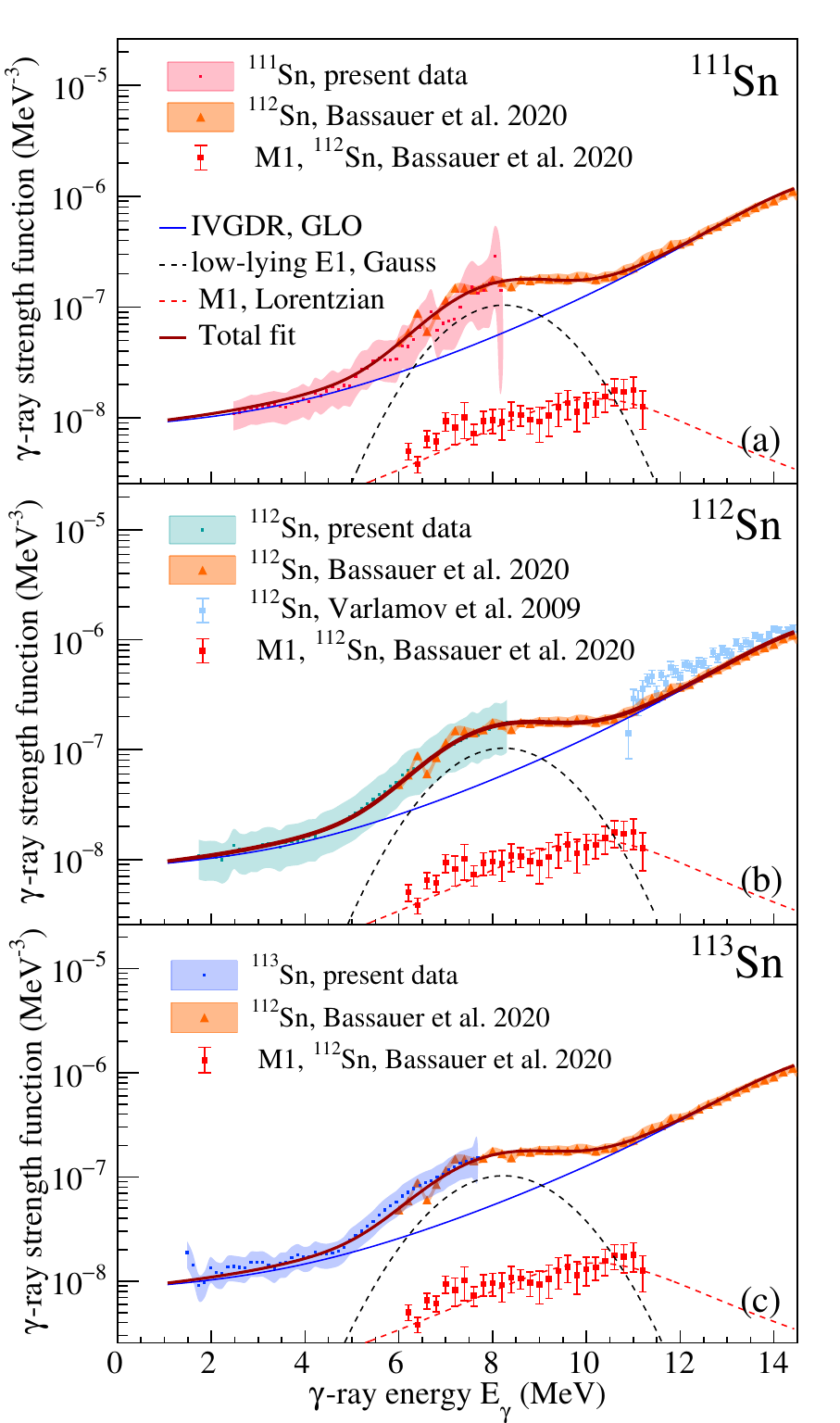}
\caption{\label{fig: GSFs fits}
The experimental GSFs for $^{111}$Sn (a), $^{112}$Sn (b), $^{113}$Sn (c) shown together with the $(p,p^{\prime})$ \cite{Bassauer2020b} and ($\gamma, n$) \cite{Varlamov2009} data for $^{112}$Sn. The total fits of the experimental data are shown as solid magenta lines and the fits of the IVGDR are marked as solid blue lines. The low-lying $E1$ and $M1$ components are shown as dashed black and red lines, respectively.
}
\end{figure}

Given the similarities in nuclear structure of $^{111-114}$Sn, we choose the ($p,p^{\prime})$ data on $^{112}$Sn to represent the region above the neutron threshold for all three isotopes, $^{111,112,113}$Sn.  
Following Refs.~\cite{Agvaanluvsan2009-2,toft2011}, the IVGDR part of the GSF is parametrized with the generalized Lorentzian function  (we exploit the same notations for all parameters):
\begin{equation}
\label{eq:15}  
    \begin{split}
        f_{E1}(E_{\gamma})=&\frac{1}{3\pi^2\hbar^2 c^2}\sigma_{E1}\Gamma_{E1}\times\\
        &\times \biggl [ E_{\gamma} \frac{\Gamma_{KMF}(E_{\gamma},T_f)}{(E_{\gamma}^2-E_{E1}^2)^2+E_{\gamma}^2\Gamma_{KMF}^2(E_{\gamma}, T_f)}+\\
        &+0.7\frac{\Gamma_{KMF}(E_{\gamma}=0,T_f)}{E_{E1}^3}\biggr ],
    \end{split}
\end{equation}
where $E_{E1}$, $\Gamma_{E1}$, $\sigma_{E1}$ are the IVGDR centroid energy, width, and cross-section, respectively. The $\Gamma_{KMF}$ parameter denotes a temperature-dependent ($T_f$) width, proposed within the Kadmenskij-Markushev-Furman  approach \cite{Kadmenskii1983}:
\begin{equation}
\label{eq:16}  
    \Gamma_{KMF}(E_{\gamma},T_f)=\frac{\Gamma_{E1}}{E_{\gamma}^2}(E_{\gamma}^2+4\pi^2T_f^2).
\end{equation}
The low-lying excess $E1$ strength superimposed on the low-energy tail of the IVGDR was found to be best described by a Gaussian-like peak:
\begin{equation}
\label{eq:17}  
    f_{low}(E_{\gamma})= C_{low}\frac{1}{\sqrt{2\pi}\sigma_{low}}\exp[-\frac{(E_{\gamma}-E_{low})^2}{2\sigma_{low}}],
\end{equation}
with $C_{low}$, $\sigma_{low}$, $E_{low}$ representing the absolute value, width, and centroid parameters, correspondingly. The experimental Oslo and Coulomb excitation data are shown together with the fitted IVGDR and the low-lying dipole strength in Fig.~\ref{fig: GSFs fits} for all three isotopes.

Since the Oslo method yields the combined $E1+M1$ dipole strength, a parametrization of the $M1$ spin-flip resonance is needed to constrain the low-lying $E1$ component. 
Previously, no experimental data on the $M1$ strength were available, and the model of Ref.~\cite{Belgya2006} was used in the earlier works \cite{Toft2010,toft2011}.
However, the new ($p,p^{\prime})$ Coulomb excitation data provide both the $E1$ and $M1$ cross sections through a multipole decomposition analysis \cite{Bassauer2020b}. The M1 cross sections can be converted to B(M1) strengths with the method described in Ref.~\cite{Birkhan2016}. The $M1$ strength appears to be quite fragmented in all of the cases \cite{Bassauer2020b}. For $^{111,112,113}$Sn, we use the $M1$ component provided by Ref.~\cite{Bassauer2020b} for $^{112}$Sn and fit it with a Lorentzian function to reproduce its overall shape:
\begin{equation}
\label{eq:18}  
    f_{M1}(E_{\gamma})= \frac{1}{3\pi^2\hbar^2c^2}\frac{\sigma_{M1}\Gamma_{M1}^2E_{\gamma}}{(E_{\gamma}^2-E_{M1}^2)^2+E_{\gamma}^2\Gamma_{M1}^2}
\end{equation}
with the maximum cross section $\sigma_{M1}$, width $\Gamma_{M1}$, and centroid $E_{M1}$. The experimental $M1$ data points are shown together with the corresponding Lorentzian fits in Fig.~\ref{fig: GSFs fits}.
%---------------------TABLE_2-------------------
\begin{table}[t]
\caption{\label{tab:table_2}Parameters used for the description of the IVGDR and the M1 strength in $^{112}$Sn.}
\begin{ruledtabular}
\begin{tabular}{lccccccc}
 Nucl. & $E_{E1}$ & $\Gamma_{E1}$ & $\sigma_{E1}$ & $T_{f}$ & $E_{M1}$ & $\Gamma_{M1}$ & $\sigma_{M1}$  \\ 
 & (MeV) & (MeV) & (mb) & (MeV) & (MeV) & (MeV) & (mb) \\
\noalign{\smallskip}\hline\noalign{\smallskip}
 $^{112}$Sn & 16.1(1)  & 5.5(3) & 266.9(95) & 0.70(5) & 10.5(4) & 4.8(5) & 1.8(2) \\
\end{tabular}
\end{ruledtabular}
\end{table}
%-----------------------------------------------

%---------------------TABLE_3-------------------
\begin{table}[]
\caption{\label{tab:table_3}Parameters used for the description of the low-lying $E1$ strengths in $^{111,112,113}$Sn, integrated low-lying $E1$ strengths, and the corresponding exhausted fractions of the TRK sum rule.}
\begin{ruledtabular}
\begin{tabular}{lccccc}
Nucl. & $E_{low}$ & $\Gamma_{low}$ & $C_{low}$ & Integrated & TRK \\ 
 & (MeV) & (MeV) & ($10^{-7}$ MeV$^{-2}$) & (MeV mb) & (\%) \\
\noalign{\smallskip}\hline\noalign{\smallskip}
$^{111}$Sn & 8.24(8)  & 1.19(6) & 3.12(23) & 29.6(15) & 1.80(10) \\
$^{112}$Sn & 8.24(9)  & 1.22(8) & 3.17(24) & 30.1(22) & 1.81(15) \\
$^{113}$Sn & 8.23(8)  & 1.25(7) & 3.21(17) & 30.5(16) & 1.82(9) \\
\end{tabular}
\end{ruledtabular}
\end{table}
%-----------------------------------------------
%---------------------------------------- FIG_8 ----------------------------------------
%---------------------------------------------------------------------------------------

The fitting approach to disentangle the $M1$ and $E1$ strengths in $^{111,112,113}$Sn is similar to that in Ref.~\cite{toft2011}. First, the $M1$ strength of $^{112}$Sn  was fitted with Eq.~(\ref{eq:18}). 
The obtained fit parameters are listed in Table \ref{tab:table_2}. Further, they were kept constant while fitting the total $E1+M1$ strength of $^{112}$Sn with the combined $f_{E1}+f_{low}+f_{M1}$ function. 
All of the IVGDR and the low-lying $E1$ strength parameters were kept free. 
Finally, the parametrization of the IVGDR for $^{112}$Sn (see Table~\ref{tab:table_2}) was applied to constrain the low-lying $E1$ strengths in $^{111}$Sn and $^{113}$Sn. 
The characteristics of all low-lying $E1$ strengths listed in Table~\ref{tab:table_3} were also used to estimate the integrated low-lying $E1$ strengths and the corresponding exhausted fractions of the classical Thomas-Reiche-Kuhn (TRK) sum rule for each isotope. 
By using the IVGDR and the $M1$ strength of $^{112}$Sn for the fit in the cases of $^{111,113}$Sn, the integrated low-lying strengths of all three isotopes yield almost the same amount of $\approx 1.8\%$ of the TRK sum rule. 
This estimate as well as the centroids $E_{low}$ are quite close to those obtained for $^{116-119,121,122}$Sn in Ref.~\cite{toft2011}, despite a slightly different approach to extract the low-lying $E1$ strength and the normalization.

The new experimental information on the GSFs of Sn isotopes below the neutron separation energy as well as the $M1$ strengths \cite{Bassauer2020b} calls for a systematic revision of all Sn isotopes studied at the OCL with a more uniform approach to the normalization of NLDs and GSFs. 
This might potentially affect the previously published parameters of the low-lying $E1$ strengths \cite{toft2011} and reveal new trends in the evolution of the low-lying strength from the lightest studied $^{111}$Sn to the heaviest $^{124}$Sn. 

%---------------------------------------- FIG_9 ----------------------------------------
\begin{figure}[t]
\includegraphics[width=1.0\columnwidth]{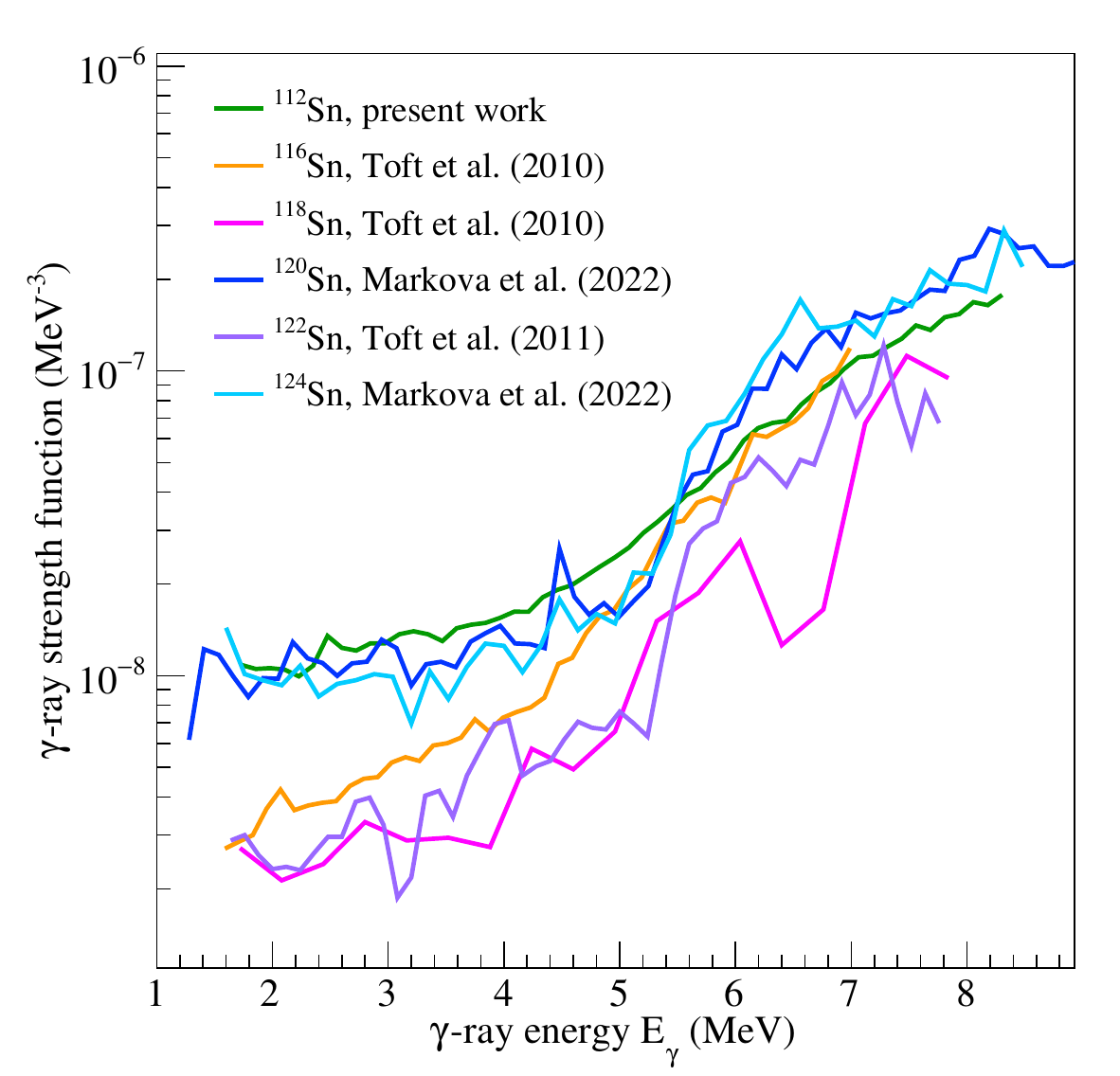}
\caption{\label{fig: GSFs even}
Comparison of experimental GSFs for $^{112}$Sn, $^{116}$Sn \cite{Toft2010}, $^{118}$Sn \cite{Toft2010}, $^{120}$Sn \cite{Markova2022}, $^{122}$Sn \cite{toft2011}, $^{124}$Sn \cite{Markova2022}. All uncertainty bands are omitted for clarity of the figure.
}
\end{figure}
%---------------------------------------------------------------------------------------

The need for a systematic re-analysis of the earlier published experiments is clearly demonstrated in Fig.~\ref{fig: GSFs even}, presenting a comparison of the GSFs for even-even Sn isotopes, namely $^{112}$Sn from the present work and already published results on $^{116,118,120,122,124}$Sn. The low-energy part of the strength is quite similar for $^{112}$Sn and the most recent results on $^{120,124}$Sn, while a clear change of the slope suggests some evolution of the strength with an increasing neutron number. The GSFs of $^{116,118,122}$Sn seem to be lower in absolute values than those of $^{112,120,124}$Sn at relatively low $E_{\gamma}$ energies. Re-normalizing these isotopes using the same models as for $^{112,120,124}$Sn and the most updated normalization information would further reveal whether this trend is due to the difference in the normalization procedures or some structural effects. For example, the spin-cutoff excitation energy dependence provided by Eq.~(\ref{eq:10}), supported by studies from Ref.~\cite{Uhrenholt13}, was chosen over other alternatives in this work as well as many other recent OCL publications (e.g. \cite{Guttormsen2022, Ingeberg2022}).  This model might potentially affect the low-energy part of the GSF, lifting it slightly up as compared to the model used in \cite{Toft2010,toft2011}. Since the time of the earlier publications (Refs.~\cite{Agvaanluvsan09-1,Agvaanluvsan2009-2,Toft2010,toft2011}) the new experimental information on s-wave neutron resonances became available for $^{116}$Sn. Even though it yields values of $\rho(S_n)$ and $\langle\Gamma_{\gamma}\rangle$ quite similar to those obtained from the systematics in \cite{Toft2010}, the systematic uncertainty band of the updated result would be considerably reduced. In addition, some issues in the normalization code that might have affected the GSF of $^{118}$Sn have been detected and fixed in the subsequent years. This appears to lead to a slightly higher GSF of $^{118}$Sn throughout the whole shown energy range. With the new neutron resonance data on $^{116}$Sn the systematics become more complete and yield new normalization parameters for $^{122}$Sn. These values are quite similar within estimated uncertainties to those in \cite{toft2011} and are not expected to change the GSF in any considerable way. However, re-visiting the energy calibration of this data set seems to yield a better fit of the NLD to the low-lying discrete states, which further shifts the updated GSF up, reaching a good agreement with the $^{120,124}$Sn GSFs as well the Coulomb excitation data. Overall, such revision of not only the even-even $^{116,118,122}$Sn, but also the odd $^{117,119,121}$Sn isotopes appears to result in a better agreement in shapes and absolute values with the recently obtained Oslo method results on $^{111-113,120,124}$Sn, the Coulomb excitation experiments \cite{Bassauer2020b}, and available ($\gamma,n$) data for all studied nuclei.

\section{\label{sec 6: Conclusion}Conclusions and outlook}

In this work, the Oslo method was used to extract the NLDs and GSFs of $^{111,112,113}$Sn from particle-$\gamma$ coincidence events obtained in the ($p,p^{\prime} \gamma$), ($p,d \gamma$), ($d,p \gamma$) reactions, respectively. 
The resulting NLDs of $^{111}$Sn and $^{113}$Sn are in good agreement with each other and the neutron evaporation data for $^{115}$Sn. 
The NLDs were used to estimate the microcanonical entropies of all three nuclei, and the entropy differences suggest an entropy of $\approx 1.5 k_B$ carried by valence neutrons in $^{111}$Sn and $^{113}$Sn. 
All three nuclei demonstrate a clear constant-temperature trend above 3 MeV in $^{111,113}$Sn and above 4.5 MeV in $^{112}$Sn, supported by the extracted microcanonical temperatures. 
 Signatures of the first neutron pair breaking  can be  seen at $\approx 2.6 - 3$ MeV in $^{111,113}$Sn and $\approx 3.6 - 4$ MeV in $^{112}$Sn. 
 Overall, the temperatures of these nuclei are quite similar to those of the neighboring $^{115,116,117}$Sn isotopes. 
 
 The GSFs extracted with the Oslo method demonstrate similar slopes for $^{111,112,113}$Sn, well in agreement within the estimated error bands with the ($p,p^{\prime})$ strengths for $^{112,114}$Sn above 6 MeV. 
 The total low-lying $E1$ strengths in these nuclei amount to $\approx 1.8 \%$ of the TRK sum rule, similar to previously published results on $^{116,117}$Sn. 
 The comparison with the new experimental information on the electric and magnetic dipole strengths from the Coulomb excitation experiments calls for a systematic revision of the earlier published $^{116-119,121,122}$Sn. This further suggests a consistent study of the evolution of the low-lying electric dipole strength with an increasing neutron number in these isotopes. A work along these lines is in progress.

\begin{acknowledgments}
The authors express their thanks to J.~C.~M\"{u}ller, P.~A.~Sobas, and J.~C.~Wikne at the Oslo Cyclotron Laboratory for operating the cyclotron and providing excellent experimental conditions. The authors are also thankful to T. W. Hagen, H. T. Nyhus, and S. J. Rose for their contribution to the experiment. This work was supported in part by the National Science Foundation under Grant No.\ OISE-1927130 (IReNA) and by the Norwegian Research Council Grants 325714 and 263030. A.~C.~L. gratefully acknowledges funding by the European Research Council through ERC-STG-2014 under Grant Agreement No.\ 637686, and from the Research Council of Norway, project number 316116. P.v.N.-C. acknowledges support by the Deutsche Forschungsgemeinschaft (DFG, German Research Foundation) under Grant No. SFB 1245 (Project ID No. 279384907).

\end{acknowledgments}

%\nocite{*}

% Produces the bibliography via BibTeX.
\bibliography{tin_2023}
\end{document}